\documentclass[graybox,vecmaths]{svmult}

\usepackage{mathptmx}       
\usepackage{helvet}         
\usepackage{courier}        
\usepackage{type1cm}        
\usepackage{makeidx}         
\usepackage{graphicx}        
\usepackage{multicol}        
\usepackage[bottom]{footmisc}

\makeindex             

\begin{document}

\title*{Wormholes and Off--Diagonal Solutions in f(R,T), Einstein and Finsler Gravity Theories}
\titlerunning{Wormholes and Off--Diagonal Solutions in Modified Gravity}
\author{Sergiu I. Vacaru}
\institute{S. Vacaru,  Alexandru Ioan Cuza University,  [Rector's office],\newline Alexandru Lapu\c sneanu street, nr. 14, Corpus R, UAIC, office 323; Ia\c si,\  700057,\  Romania;\newline
\email{sergiu.vacaru@uaic.ro, Sergiu.Vacaru@gmail.com}}
%
%
\maketitle
\abstract{The aims of this work are 1) to sketch a proof  that there are such parameterizations of the local frame and canonical connection structures when the gravitational field equations in f(R,T)--modified gravity, MG, can be integrated in generic off--diagonal forms with metrics depending on all spacetime coordinates and 2) to provide some examples of exact solutions.}

\section{Nonholonomic Deformations in Modified Gravity Theories}

\label{s2}
We study gravity theories formulated on a  spacetime manifold $\mathbf{V}, dim \mathbf{V}= n \geq 4$  (for Finsler models, on tangent bundle $T\mathbf{V}$)   endowed with metric,  $\mathbf{g}$, and compatible linear connection $\mathbf{D} $, structures, $\mathbf{Dg}=0$,  see details in Refs. \cite{vadm1,vfinsl1,vfinsl2,vfinsl3}. Our goal is to prove   that there are such local frame and canonical connection structures when the gravitational field equations in $f(R,T)$--modified gravity, MG, see reviews  \cite{odints1,odints2,odints3,stv}, can be integrated in generic off--diagonal forms with metrics depending on all spacetime coordinates. We  provide explicit  examples when generalized solutions in MG can be equivalently modelled as  effective Einstein spaces and determine  deformations of wormhole spacetimes in general relativity (GR).

\vskip3pt

 {\bf 1.1\ Geometric Preliminaries}:\ We  consider a conventional horizontal (h) and vertical (v) splitting of the tangent space $T\mathbf{V,}$ when a non--integrable (equivalently, nonholonomic, or anholonomic) distribution  $\mathbf{N}:\ T\mathbf{V}=h\mathbf{V}\oplus v\mathbf{V}$ (for Finsler theories, $\mathbf{N}:\ TT\mathbf{V}=hT\mathbf{V}\oplus vT\mathbf{V}$). Locally, such a h--v--splitting is determined by a set of coefficients $\mathbf{N}=\{N_{i}^{a}(x,y)\}$ and  coordinates  parameterized:\ $u=(x,y)$, $u^{\mu }=(x^{i},y^{a}),$ where the h--(v--)indices run  $i,j,...=1,2,...,n$ ($a,b,...=n+1,...,n+n$).  There are N--adapted  frames $\mathbf{e}_{\nu }=(\mathbf{e}_{i}, e_{a})$,  $\mathbf{e}^{\mu}=(e^{i},\mathbf{e}^{a}),$
\begin{equation}
\mathbf{e}_{i} = \partial /\partial x^{i}-\ N_{i}^{a}(u)\partial /\partial
y^{a},\ e_{a}=\partial _{a}=\partial /\partial y^{a},\  
 e^{i} = dx^{i},\ \mathbf{e}^{a}=dy^{a}+\ N_{i}^{a}(u)dx^{i}, \label{nadif}
\end{equation}%
which satisfy the conditions $\lbrack \mathbf{e}_{\alpha },\mathbf{e}_{\beta }]=\mathbf{e}_{\alpha }%
\mathbf{e}_{\beta }-\mathbf{e}_{\beta }\mathbf{e}_{\alpha }=W_{\alpha \beta
}^{\gamma }\mathbf{e}_{\gamma }$,  with anholonomy coefficients
 $W_{ia}^{b}=\partial _{a}N_{i}^{b},W_{ji}^{a}=\Omega _{ij}^{a}=\mathbf{e}%
_{j}\left( N_{i}^{a}\right) -\mathbf{e}_{i}(N_{j}^{a})$.

On a nonholonomic manifold $(\mathbf{V,N})$, and/or nonholonomic bundle $(T\mathbf{V,N})$, we can represent any data $(\mathbf{g,D})$ in N--adapted form (preserving under parallel transport a chosen  h-v--splitting) parameterized as: 1) a \textit{distinguished metric, d--metric,}
\begin{equation}
\mathbf{g}=g_{\alpha }(u)\mathbf{e}^{\alpha }\otimes \mathbf{e}^{\beta
}=g_{i}(x)dx^{i}\otimes dx^{i}+g_{a}(x,y)\mathbf{e}^{a}\otimes \mathbf{e}%
^{a}.  \label{dm1}
\end{equation}
and 2) a  \textit{distinguished connection, d--connection,} $\mathbf{D}=(hD,vD)$.

Any d--connection is characterized by  d--torsion,   nonmetricity,  and d--curvature structures:
 $\mathcal{T}(\mathbf{X,Y}) :=\mathbf{D}_{\mathbf{X}}\mathbf{Y}-\mathbf{D}_{%
\mathbf{Y}}\mathbf{X}-[\mathbf{X,Y}],$ $\mathcal{Q}(\mathbf{X}):=\mathbf{D}_{%
\mathbf{X}}\mathbf{g,}$  $\mathcal{R}(\mathbf{X,Y}) :=\mathbf{D}_{\mathbf{X}}\mathbf{D}_{\mathbf{Y}}-%
\mathbf{D}_{\mathbf{Y}}\mathbf{D}_{\mathbf{X}}-\mathbf{D}_{\mathbf{[X,Y]}}$, where $\mathbf{X,Y}\in T\mathbf{V}$ (or $\in TT\mathbf{V}$, in Finsler like theories).

There are two "preferred" linear connections which can be defined for the same data $(\mathbf{g,N})$: 1) the \textit{canonical d--connection} $%
\widehat{\mathbf{D}}$ uniquely determined by the conditions that it is metric compatible, $\widehat{\mathbf{D}}\mathbf{g=0,}$ and  with zero h--torsion, $h\widehat{%
\mathcal{T}}=\{\widehat{T}_{\ jk}^{i}\}=0,$ and zero v--torsion, $v\widehat{\mathcal{T}}=\{\widehat{T}_{\ bc}^{a}\}=0$;  2) the Levi--Civita (LC) connection,  $\nabla$,
when $\mathcal{T}=0 $ and $\mathcal{Q}=0$, if $\mathbf{D}\rightarrow \nabla$.  Such linear connections are related via a canonical distortion relation  $\widehat{\mathbf{D}}=\nabla +\widehat{\mathbf{Z}}$. We can work equivalently on $\mathbf{V}$ and $T\mathbf{V}$  using  both linear connections.
For any  data $(\mathbf{g,N,}\widehat{\mathbf{D}})$, we can define and compute in standard form, respectively, the Riemann, $\widehat{\mathcal{R}}=\mathbf{\{}%
\widehat{\mathbf{R}}_{\ \beta \gamma \delta }^{\alpha }\},$ and the Ricci, $\widehat{\mathcal{R}}ic=\{\widehat{\mathbf{R}}_{\alpha \beta }:=\widehat{\mathbf{R}}_{\ \alpha \beta \gamma }^{\gamma }\}$ d--tensors; for
  $\widehat{R}:=\mathbf{g}^{\alpha \beta }\widehat{\mathbf{R}}_{\alpha \beta
}$, we can introduce  $\widehat{\mathbf{E}}_{\alpha \beta }:= \widehat{\mathbf{R}}_{\alpha
\beta }-\frac{1}{2}\mathbf{g}_{\alpha \beta }\ \widehat{R}$.

\vskip3pt

 {\bf 1.2\ Nonholonomically Modified Gravity}:\ We study  theories with  action%
\begin{equation}
S=\frac{1}{16\pi }\int \delta u^{n+n}\sqrt{|\mathbf{g}_{\alpha \beta }|}[f(%
\widehat{R},T)+\ ^{m}L],  \label{act}
\end{equation}%
 generalizing the so--called modified $f(R,T)$ gravity \cite{odints1,odints2,odints3} to the case of d--connection $\widehat{\mathbf{D}}$, which can be considered for (pseudo) Riemannian spaces (as an "auxiliary" one) \cite{vadm1}, for Ho\v rava--Lifshits type modifications \cite{vfinsl2,vhlquant} and on (non) commutative  Finsler spaces \cite{vfinsl1,vfinsl3,stv}. In (\ref{act}), $T$ is the trace of the stress--energy momentum tensor constructed for the matter fields Lagrangian $\ ^{m}L$. It is possible to elaborate a N--adapted variational formalism   for a large class of models with perfect fluid matter with $\ ^{m}L=-p$, for pressure $p$, and assuming that
 $f(\widehat{R},T)=\ ^{1}f(\widehat{R})+\ ^{2}f(T)$, where  $\ ^{1}F(\widehat{R}):=\partial \ ^{1}f(\widehat{R})/\partial
\widehat{R}$ and $\ ^{2}F(T):=\partial \ ^{2}f(T)/\partial T$. We obtain a model of MG with effective Einstein equations,  $\widehat{\mathbf{E}}_{\alpha \beta }={\Upsilon }_{\beta \delta }$,
  for source
 ${\Upsilon }_{\beta \delta }=\ ^{ef}\eta \ G\ \mathbf{T}_{\beta \delta
}+\ ^{ef}\mathbf{T}_{\beta \delta }$, where   $\ ^{ef}\eta =[1+\ ^{2}F/8\pi ]/\
^{1}F $ is the  effective
polarization of cosmological constant  $G$, $\mathbf{T}_{\beta \delta
}$ is the usual energy--momentum tensor for matter fields and the $f$--modification of the energy--momentum tensor results in
  $\ ^{ef}\mathbf{T}_{\beta \delta }=[\frac{1}{2}(\ ^{1}f-\ ^{1}F\ \widehat{R}%
+2p\ ^{2}F+\ ^{2}f)\mathbf{g}_{\beta \delta }-(\mathbf{g}_{\beta \delta }\
\widehat{\mathbf{D}}_{\alpha }\widehat{\mathbf{D}}^{\alpha }-\widehat{%
\mathbf{D}}_{\beta }\widehat{\mathbf{D}}_{\delta })\ ^{1}F]/\ ^{1}F$.

The effective Einstein equations 
  decouple  for parameterizations of metrics (\ref{dm1}) when the coefficients $N_{i}^{a}(u)$ in   (\ref{nadif}) are such way prescribed that the corresponding  nonholonomic constraints result in $\widehat{\mathbf{D}}$ with $\widehat{R}=const$ and ${\Upsilon }_{~\delta
}^{\beta }=(\Lambda +\lambda )\mathbf{\delta }_{~\delta }^{\beta }$ for an effective cosmological constant
$\Lambda $ for modified gravity and $\lambda$ for a possible cosmological constant in GR. This results in  $\widehat{\mathbf{D}}_{\delta }\ ^{1}F_{\mid \Upsilon =\Lambda +\lambda}=0$, see details in \cite{vadm1,vfinsl1,vfinsl2}.

\section{Ellipsoid, Solitonic \& Toroid Deformations of Wormholes}

 The general stationary ansatz for off--diagonal solutions   is
{\small
\begin{eqnarray}
\mathbf{ds}^{2} &=&e^{\tilde{\psi}(\widetilde{\xi },\theta )}(d\widetilde{%
\xi }^{2}+\ d\vartheta ^{2})+  \frac{\lbrack \partial _{\varphi }\varpi (\widetilde{\xi },\vartheta
,\varphi )]^{2}}{ \Lambda + \lambda }
\left( 1+\varepsilon \frac{\partial _{\varphi }[\chi _{4}(\widetilde{\xi }%
,\vartheta ,\varphi )\varpi (\widetilde{\xi },\vartheta ,\varphi )]}{%
\partial _{\varphi }\varpi (\widetilde{\xi },\vartheta ,\varphi )}\right) \nonumber \\
&& r^{2}(\widetilde{\xi })\sin ^{2}\theta (\widetilde{\xi },\vartheta )(\delta
\varphi )^{2}   -\frac{e^{2\varpi (\widetilde{\xi },\vartheta ,\varphi )]}}{| \Lambda + \lambda | }[1+\varepsilon \chi _{4}(%
\widetilde{\xi },\vartheta ,\varphi )]e^{2B(\widetilde{\xi })}(\delta t)^{2},
 \label{offdwans1} \\
\delta \varphi &=&d\varphi +\partial _{\widetilde{\xi }}[\ ^{\eta }%
\widetilde{A}(\widetilde{\xi },\vartheta ,\varphi )+\varepsilon \overline{A}(%
\widetilde{\xi },\vartheta ,\varphi )]d\widetilde{\xi }+\partial _{\vartheta
}[\ ^{\eta }\widetilde{A}(\widetilde{\xi },\vartheta ,\varphi )+\varepsilon
\overline{A}(\widetilde{\xi },\vartheta ,\varphi )]d\vartheta ,\   \nonumber  \\
\delta t &=&dt+\partial _{\widetilde{\xi }}[\ ^{\eta }n(\widetilde{\xi }%
,\vartheta )+\varepsilon \partial _{i}\overline{n}(\widetilde{\xi }%
,\vartheta )]~d\widetilde{\xi }+\partial _{\vartheta }[\ ^{\eta }n(%
\widetilde{\xi },\vartheta )+\varepsilon \partial _{i}\overline{n}(%
\widetilde{\xi },\vartheta )]~d\vartheta ,  \nonumber
\end{eqnarray}%
} where $\ \widetilde{\xi }=\int dr/\sqrt{|1-b(r)/r|}$ for  $b(r),B(\widetilde{\xi
}) $  determined by a wormhole metric in GR. For 4--d theories, we consider  $x^i=(\widetilde{\xi },\theta)$ and $y^a=(\varphi,t)$.

\vskip3pt

{\bf 2.1\ Rotoid --configurations} with a small parameter (eccentricity) $\varepsilon $ are "extracted" from (\ref{offdwans1})
if we take for the $f$--deformations
 $\chi _{4}=\overline{\chi }_{4}(r,\varphi ):=\frac{2\overline{M}(r)}{r}$\newline $\left(
1-\frac{2\overline{M}(r)}{r}\right) ^{-1}\underline{\zeta }\sin (\omega
_{0}\varphi +\varphi _{0})$,
 for $r$ considered as a function $r(\ \widetilde{\xi })$.
Let us define
{\small
\begin{eqnarray}
&& h_{3} =\widetilde{\eta }_{3}(\widetilde{\xi },\vartheta ,\varphi )[
1+\varepsilon \chi _{3}(\widetilde{\xi },\vartheta ,\varphi )]\ ^{0}%
\overline{h}_{3}(\widetilde{\xi },\vartheta ),\
h_{4} = \widetilde{\eta }_{4}(\widetilde{\xi },\vartheta ,\varphi )[
1+\varepsilon \overline{\chi }_{4}(\widetilde{\xi },\varphi )] \ ^{0}%
\overline{h}_{4}(\widetilde{\xi }), \mbox{for} \nonumber \\
&& \ ^{0}\overline{h}_{3}=r^{2}(\widetilde{\xi })\sin ^{2}\theta (%
\widetilde{\xi },\vartheta ), \ ^{0}\overline{h}_{4}=q(\widetilde{\xi }),\
\widetilde{\eta }_{3}=\frac{[\partial _{\varphi }\varpi (\widetilde{\xi }%
,\vartheta ,\varphi )]^{2}}{ \Lambda +\lambda  },\widetilde{\eta }_{4}=\frac{e^{2\varpi (\widetilde{\xi }%
,\vartheta ,\varphi )]}}{| (\Lambda + \lambda) |q(\widetilde{\xi })}e^{2B(\widetilde{\xi })},  \label{polfew}
\end{eqnarray}%
}
where $e^{2B(\widetilde{\xi })}\rightarrow q(\widetilde{\xi })$ if $\widetilde{\xi }\rightarrow \xi .$ For a prescribed  $\widetilde{%
\varpi }(\widetilde{\xi },\vartheta ,\varphi ),$ we  compute
 $\widetilde{\chi }_{3}=\chi _{3}(\widetilde{\xi },\vartheta ,\varphi ) =\partial _{\varphi }[\overline{\chi }_{4}\widetilde{\varpi }%
]/\partial _{\varphi }\widetilde{\varpi }$,   $\overline{w}_{i}=\frac{\partial _{i}(\ r(\widetilde{\xi })\sin \theta (%
\widetilde{\xi },\vartheta )\sqrt{|q(\widetilde{\xi })|}\partial _{\varphi }[%
\overline{\chi }_{4}\varpi ])}{e^{\varpi }r(\widetilde{\xi })\sin \theta (%
\widetilde{\xi },\vartheta )\sqrt{|q(\widetilde{\xi })|}\partial _{\varphi
}\varpi }=\partial _{i}\overline{A}(\widetilde{\xi },\vartheta ,\varphi ),$
for $x^{i}=(\widetilde{\xi },\vartheta ).$ We model an ellipsoid configuration with $\ r_{+}(\widetilde{\xi }_{+})\simeq \frac{2\ \overline{M}(\ \widetilde{\xi }_{+})}{%
1+\varepsilon \underline{\zeta }\sin (\omega _{0}\varphi +\varphi _{0})}$,
for  constants $\underline{%
\zeta },\omega _{0}, \varphi _{0}$ and eccentricity $\varepsilon .$ We obtain{\small
\begin{eqnarray}
\mathbf{ds}^{2} &=&e^{\tilde{\psi}(\widetilde{\xi },\theta )}(d\widetilde{%
\xi }^{2}+\ d\vartheta ^{2})+   \frac{[\partial _{\varphi }\widetilde{\varpi }]^{2}}{\Lambda +
\lambda }(1+\varepsilon \frac{\partial _{\varphi }[\overline{\chi }_{4}%
\widetilde{\varpi }]}{\partial _{\varphi }\widetilde{\varpi }})\ ^{0}%
\overline{h}_{3}[d\varphi +\partial _{\widetilde{\xi }}(\ ^{\eta }\widetilde{%
A}+\varepsilon \overline{A})d\widetilde{\xi }+\partial _{\vartheta }(\
^{\eta }\widetilde{A} + \nonumber  \\ && \varepsilon \overline{A})d\vartheta ]^{2}
-\frac{e^{2\widetilde{\varpi }}}{|\Lambda +  \lambda |}[1+\varepsilon
\overline{\chi }_{4}(\widetilde{\xi },\varphi )]e^{2B(\widetilde{\xi }%
)}[dt+\partial _{\widetilde{\xi }}(\ ^{\eta }n+\varepsilon \overline{n})~d%
\widetilde{\xi }+\partial _{\vartheta }(\ ^{\eta }n+\varepsilon \overline{n}%
)~d\vartheta ]^{2}.  \label{ellipswh}
\end{eqnarray}%
}If the generating functions $\widetilde{\varpi }$ and effective sources   are such way chosen that the polarization functions
(\ref{polfew}) can be approximated $\widetilde{\eta }_{a}\simeq 1$ and $%
^{\eta }\widetilde{A}$ and $\ ^{\eta }n$ are "almost constant",  the metric (\ref{ellipswh}) mimics
small rotoid wormhole like configurations.
\vskip3pt

{\bf 2.2\ Solitonic waves, wormholes and black ellipsoids:}\  An interesting class of off--diagonal solutions depending on all spacetime
coordinates can be constructed by designing a configuration when a
1--solitonic wave preserves an ellipsoidal wormhole configuration. Such a
spacetime metric can be written in the form
{\small
\begin{equation}
\mathbf{ds}^{2} = e^{\tilde{\psi}(x^{i})}(d\widetilde{\xi }^{2}+\
d\vartheta ^{2})+\omega ^{2} \left[ \widetilde{\eta }_{3}(1+\varepsilon \frac{\partial _{\varphi }[%
\overline{\chi }_{4}\widetilde{\varpi }]}{\partial _{\varphi }\widetilde{%
\varpi }})\ ^{0}\overline{h}_{3}(\delta \varphi )^{2}-\widetilde{\eta }%
_{4}[1+\varepsilon \overline{\chi }_{4}(\widetilde{\xi },\varphi )]\ ^{0}%
\overline{h}_{4}(\delta t)^{2}\right],  \label{solitwh}
\end{equation}
}
for $\delta \varphi = d\varphi +\partial _{i}(\ ^{\eta }\widetilde{A}%
+\varepsilon \overline{A})dx^{i},\delta t=dt+~_{1}n_{i}(\widetilde{\xi }%
,\vartheta )dx^{i}$,    $x^{i}=(\widetilde{\xi },\vartheta )$ and $y^{a}=(\varphi ,t).$ The
 factor
 $\omega (\widetilde{\xi },t)=4\arctan e^{m\gamma (\widetilde{\xi }-vt)+m_{0}}$,
where $\gamma ^{2}=(1-v^{2})^{-1}$ and constants $m,m_{0},v,$ defines a
1--soliton for the sine--Gordon equation,\
 $\frac{\partial ^{2}\omega }{\partial t^{2}}-\frac{\partial ^{2}\omega }{%
\partial \widetilde{\xi }^{2}}+\sin \omega =0$.

For $\omega =1,$ the metrics (\ref{solitwh}) are of type (\ref{ellipswh}). A
nontrivial value $\omega $ depends on the time like coordinate $t$ and has
to be constrained to certain conditions which do not change the Ricci d--tensor,  which can be written
for $~_{1}n_{2}=0$ and $~_{1}n_{1}=const$ in the form $\frac{\partial \omega
}{\partial \widetilde{\xi }}-~_{1}n_{1}\frac{\partial \omega }{\partial t}=0$. A gravitational solitonic wave  propagates
self--consistently in a rotoid wormhole background with $_{1}n_{1}=v$ which
solve both the sine--Gordon and constraint equations. Re--defining the
system of coordinates with $x^{1}=\widetilde{\xi }$ and $x^{2}=\theta ,$ we
can transform any  $~_{1}n_{i}(\widetilde{\xi },\theta )$ into
necessary $_{1}n_{1}=v$ and $_{1}n_{2}=0.$

\vskip3pt

{\bf 2.3\ Ringed wormholes:}\ We can generate  a rotoid wormhole plus a torus,  {\small
 $$\mathbf{ds}^{2} =e^{\tilde{\psi}(x^{i})}(d\widetilde{\xi }^{2}+\
d\vartheta ^{2})+\widetilde{\eta }_{3}(1+\varepsilon \frac{\partial
_{\varphi }[\overline{\chi }_{4}\widetilde{\varpi }]}{\partial _{\varphi }%
\widetilde{\varpi }})\ ^{0}\overline{h}_{3}(\delta \varphi )^{2}  -f \widetilde{\eta }%
_{4}[1+\varepsilon \overline{\chi }_{4}(\widetilde{\xi },\varphi )]\ ^{0}%
\overline{h}_{4}(\delta t)^{2},$$
}
for $\delta \varphi = d\varphi +\partial _{i}(\ ^{\eta }\widetilde{A}%
+\varepsilon \overline{A})dx^{i},\delta t=dt+~_{1}n_{i}(\widetilde{\xi }%
,\vartheta )dx^{i}$, when $x^{i}=(\widetilde{\xi },\vartheta )$ and $y^{a}=(\varphi ,t),$ where
the function $f(\widetilde{\xi },\vartheta ,\varphi )$  can be rewritten equivalently in
Cartesian coordinates,
 $f(\widetilde{x},\widetilde{y},\widetilde{z})=\left( R_{0}-\sqrt{\widetilde{x}%
^{2}+\widetilde{y}^{2}}\right) ^{2}+\widetilde{z}^{2}-a_{0}$,
for $a_{0}<a,R_{0}<r_{0}$. We get a ring around the wormhole throat
 (we  argue that we obtain well--defined wormholes in the limit $\varepsilon
\rightarrow 0$ and for corresponding approximations $\widetilde{\eta }%
_{a}\simeq 1$ and $^{\eta }\widetilde{A}$ and $\ ^{\eta }n$ to be almost
constant). The ring configuration is stated by the condition $\ f=0.$  In above
formulas, $R_{0}$ is the distance from the center of the tube to the center
of the torus/ring and $a_{0}$ is the radius of the tube. If the wormhole objects exist, the variants ringed by a torus may be stable
for certain nonholonomic geometry and exotic matter configurations.

\vskip3pt

{\bf Acknowledgments:}  Research is  supported by  IDEI, PN-II-ID-PCE-2011-3-0256.

\end{document}